\documentclass[11pt]{iopart}
\usepackage{graphicx,epsfig,iopams}
\begin{document}

\title[Colloid under AC Fields]{Dielectric Response of Nanoscopic Spherical Colloids in Alternating Electric Fields:\\ A Dissipative Particle Dynamics Simulation}    

\author{Jiajia Zhou and Friederike Schmid}
\address{Institut f\"ur Physik, Johannes Gutenberg-Universit\"at Mainz,\\
 Staudingerweg 7, D-55099 Mainz, Germany} 
\eads{\mailto{zhou@uni-mainz.de} and \mailto{friederike.schmid@uni-mainz.de}}

\begin{abstract} 
We study the response of single nanosized spherical colloids in electrolyte
solution to an alternating electric field (AC field) by computer simulations.
We use a coarse-grained mesoscopic simulation approach that accounts in full
for hydrodynamic and electrostatic interactions as well as for thermal
fluctuations.  The solvent is modeled as a fluid of single Dissipative Particle
Dynamics (DPD) beads, and the colloidal particle is modeled as a rigid body
made of DPD beads.  We compute the mobility and the polarizability of a single
colloid and investigate systematically the effect of amplitude and frequency of
the AC-fields.  Even though the thickness of the Debye layer is not ``thin''
compared to the radius of the colloid, and the thermal fluctuations are
significant, the results are in good agreement with the theoretical prediction
of the Maxwell-Wagner-O'Konski theory, especially for uncharged colloids.
\end{abstract}


\submitto{\JPCM}

\maketitle

\section{Introduction}
\label{sec:introduction}

Alternating electric fields (AC fields) can be used to manipulate single
colloid or colloidal suspensions.  This provides an attractive tool for
controlling the position of individual colloid or the structure of colloidal
assemblies.  Since the colloids respond to external fields on relatively short
time scales and in an often fully reversible way, the approach provides high
flexibility in designing optimal AC pulses for a given task.

One particularly prominent example is dielectrophoresis.  In dielectrophoresis,
one exploits the fact that external electric fields polarize the colloids,
which are then in turn driven along the direction of the field gradient.  The
time-averaged force acting on a particle in the AC-field ${\bf E} \exp(i\omega
t)$ is \begin{equation} {\bf F}_{DEP} \sim {\rm Re} \{ \alpha(\omega) \} \nabla
|{\bf E}|^2, \end{equation} where $\alpha(\omega)$ is the complex
polarizability of the particle.  The dielectrophoretic effect can be used for
trapping colloidal suspensions in ``electric bottles'' \cite{Sullivan2006,
Leunissen2008}, but also for selectively trapping single colloid and separating
colloids or viruses by size and shape \cite{Green1997, Morgan1999}.  It has
thus many potential applications in nano- and biotechnology.

The crucial quantity in dielectrophoresis is the polarizability
$\alpha(\omega)$.  Assuming that the external field has a form ${\bf E}
\exp(i\omega t)$, the effective dipole moment of a spherical particle can be
written as

\begin{equation}
  \label{eq:mosotti}
  {\bf p} = \alpha(\omega) {\bf E} = 4\pi \epsilon_m K(\omega) R^3 \, {\bf E},
\end{equation}
where $\epsilon_m$ is the permittivity of the surrounding medium, $K(\omega)$
is the Clausius-Mosotti factor and $R$ is the radius of the particle.  For
uncharged particles, the polarizability is governed by the permittivity and
conductivities of the particle and the medium.  For charged particles in salt
solution, the electrical double layer surrounding the particle contributes
significantly to the polarizability, and the dielectrophoretic response results
from an intricate interplay of dipole reorientation, ion mobility,
electrostatics, and hydrodynamics. The dielectric response of colloids to AC
fields has attracted abiding interest for decades, but it is still not fully
understood, especially in cases where the colloids have nanometer size and 
the electric double layer is thick, since most theoretical studies neglect
thermal fluctuations and assume that the electric double layer is thin compared
to the colloidal radius.

Several processes, on time scales spanning several orders of magnitude,
contribute to the dielectric response of charged colloids in salt solutions.
The fastest event following the sudden application of a constant electric field
is the interface polarization induced by the local dielectric contrast between
water and the colloid material.  This happens on the GHz scale (the relaxation
time of water is 17 GHz), which is instantaneous for the purpose of the present
discussion.  Likewise, the time for hydrodynamic flow to build up is also
almost instantaneous \cite{Shilov2000}.  On the MHz scale (1-10 MHz), ionic
migration in the electrical double layer sets in and dominates over the
molecular dipole reorientation.  In this regime, the main contribution to the
induced dipole moments stems from the conductivity mismatch between the
particle and the solvent due to the presence of free counterions in the solvent
(Maxwell-Wagner relaxation).  Finally, on long time scales up to 100Hz, a salt
concentration gradient builds up along the colloid and the thickness of the
double layer varies accordingly, leading to an additional source of polarization
($\alpha$-polarization). 

Various approaches have been proposed to explain the dielectric response of
colloidal particles.  In the high-frequency regime, the Maxwell-Wagner
mechanism of dielectric dispersion \cite{Maxwell1954, Wagner1914} has been
widely discussed, which only depends on the bulk properties of the solution and
the colloid.  O'Konski later introduced a surface conductivity term to account
for the contribution of the electric double layer \cite{OKonski1960}. The
Maxwell-Wagner-O'Konski (MWO) theory has also been extended to ellipsoidal
particles \cite{Saville2000}. Theories for the low-frequency region have been
developed in the Ukraine school \cite{DukhinShilov} based on the standard
electrokinetic model \cite{RSS}. Most studies rely on the assumption that the
electrical double layer is much thinner than the radius of the particle. For
situations that involve thick electrical double layers and the whole frequency
spectrum, numerical methods to solve the electrokinetic equations have been
implemented \cite{DeLacey1981, Hill2003, Zhaohui2009}. In the standard
electrokinetic model, ions are treated as point particles, the size effect is
neglected, and a mean field approach is taken, i.e., thermal fluctuations are
neglected. Moreover, linear response theory is employed in the numerical
calculations, but experiments have suggested that the dielectric response
becomes nonlinear already for moderate field strengths \cite{ZhuJT2007}. 

Molecular simulations can shed light on dielectrophoretic phenomena in a
well-defined model system, but are still scarce \cite{Salonen2005,
Salonen2007}.  Such studies are numerically challenging, because two different
types of long-range interactions are involved: the electrostatic and the
hydrodynamic interactions.  In recent years, a number of coarse-grained
simulation approaches have been developed to address this class of problem.
The idea is to couple explicit charges (ions, colloids) with a mesoscopic model
for Navier-Stokes fluids. On the side of the fluid model, different approaches
exist in the literature that range from Lattice Boltzmann (LB) methods
\cite{Lobaskin2004, Chatterji2005}, Multi-Particle Collision Dynamics (MPCD)
\cite{Malevanets1999, Gompper2009}, to Dissipative Particle Dynamics (DPD)
\cite{Koelman1993, Espanol1995, Groot1997}.  In this paper, we use the DPD
method.  DPD is a coarse-grained simulation method which is momentum-conserving
and Galilean invariant.  Since it is a particle-based method, charges can be
introduced quite naturally.  A recent comparative study \cite{Smiatek2009}
indicated that the Coulomb interaction is the most time-consuming part in the
simulation.  The computational costs of different methods for modeling the
fluid thus becomes comparable at intermediate or high salt concentrations. 

In the present paper, we use DPD simulation to study the dielectric response of
a single spherical colloid subject to AC fields, accounting in full for the
hydrodynamic and electrostatic interactions.  The remainder of this article is
organized as follows: In section \ref{sec:model}, we give a brief introduction
of our simulation model.  We present the simulation results on the
polarizability and effects of systematically varying the properties of the
external AC fields in section \ref{sec:results}.  Finally, section
\ref{sec:summary} concludes with a brief summary.

\section{Model}
\label{sec:model}

Our simulation system has three components: The solvent, the colloidal
particle and the microions. The solvent is modeled as a fluid of DPD beads
without conservative interactions, i.e., we only implement the DPD dissipative
and stochastic interactions without conservative forces \cite{Soddemann2003}.
In the following, physical quantities will be reported in a model unit system
of $\sigma$ (length) $m$ (mass), $\varepsilon$ (energy), and $e$ (elementary
charge unit). In these units, the temperature of the system is
$k_BT=1.0\varepsilon$, the number density of the fluid is $3.0 \sigma^{-3}$,
and each solvent bead has a mass $1 m$. The DPD friction coefficient is set to
$\gamma_{DPD}=5.0 \sqrt{m\varepsilon}/\sigma$ and the cutoff radius is $r_c =
1.0 \sigma$.  The shear viscosity of the fluid was evaluated by analyzing pure
Poiseuille flows in a microchannel by a method described in Ref.
\cite{Smiatek2008}, and was found to be $\eta_s=1.23 \pm 0.01
\sqrt{m\varepsilon}/\sigma^2$, in good agreement with Ref.\ \cite{Smiatek2009}.
The diffusion constant of single solvent beads is measured to be 
$D_S=0.66 \pm 0.02\,\sigma\sqrt{\varepsilon/m}$.

The colloidal particle is represented by a large sphere which has repulsive
Weeks-Chandler-Anderson (WCA)\cite{WCA} type conservative interactions with
the fluid particles,
\begin{equation}
  \label{eq:WCA}
  V_{WCA}(r) = \left\{ \begin{array}{ll}
      4\varepsilon \left[ ( \frac{\sigma}{r-r_0} )^{12} 
           - ( \frac{\sigma}{r-r_0} )^6 + \frac{1}{4} \right] \quad & \mbox{for } \quad r<r_c \\
       0 & \mbox{otherwise}
      \end{array} \right.
\end{equation}
where the cutoff radius is set at the potential minimum $r_c=r_0 + 2^{1/6}
\sigma$.  The radius of the colloid is $R=r_0+\sigma=3.0 \sigma$. To implement
no-slip boundary conditions at the surface, a set of $N_s$ DPD interaction
sites is distributed evenly on the surface, with positions that are fixed with
respect to the colloid center. These sites interact with the solvent beads through
the DPD dissipative and stochastic interactions, using the same DPD friction
constant as in the fluid, but twice the cutoff range.  (The DPD cutoff has to
be larger than the range of the WCA interaction in order to ensure that a
sufficient number of fluid beads will interact with the DPD surface sites.) The
total force exerted on the colloid is given by the sum over all DPD
interactions with surface sites, plus the conservative excluded volume
interaction,
\begin{equation}
  {\bf F}_C = \sum^{N_s}_{i=1} {\bf F}^{(s)}_i({\bf r}_i) + {\bf F}^{(WCA)}.
\end{equation}
Here ${\bf r}_i$ denotes the position of $i$-th surface sites. 
Similarly, the torque exerted on the colloid can be written as
\begin{equation}
  {\bf T}_C = \sum^{N_s}_{i=1} {\bf F}^{(s)}_i({\bf r}_i) 
   \times ({\bf r}_i - {\bf r}_{cm}),
\end{equation}
where ${\bf r}_{cm}$ is the position vector of the colloid's center-of-mass.
Here the excluded volume interaction does not contribute since it points
towards the colloid center.  The force and the torque are used to update the
position and velocity of the colloid in a time step using the Velocity-Verlet
algorithm.  The mass of the colloidal particle is $M=100 m$ and the moment of
inertia is $I=360 m\sigma^2$, corresponding to a uniformly distributed mass. 

In addition, a total charge $Q$ may be assigned to the center of the colloidal
particle. To keep the whole system charge-neutral, a corresponding amount of
counterions is then added to the solution. Salt microions may also be added 
as pairs of monovalent positively and negatively charged beads.  We only
consider the monovalent case where counterions or salt ions carry a single
elementary charge $\pm 1 e$. Charges interact by Coulomb interactions, and the
Bjerrum length $l_B=e^2/(4\pi \epsilon_m k_BT)$ of the fluid is set to $1.0
\sigma$.  Apart from the electrostatic interaction with other charged
particles, the ions have a short-range repulsive WCA interaction with the
solvent beads and each other, which has the form of Eq.\ (\ref{eq:WCA}) 
with $r_{0}=0$. The conservative interaction with the colloid is also given
by Eq.\ (\ref{eq:WCA}) with $r_{0}+\sigma = 3.0\sigma$.

\begin{figure}[htbp]
  \centerline{\includegraphics[width=0.4\columnwidth]{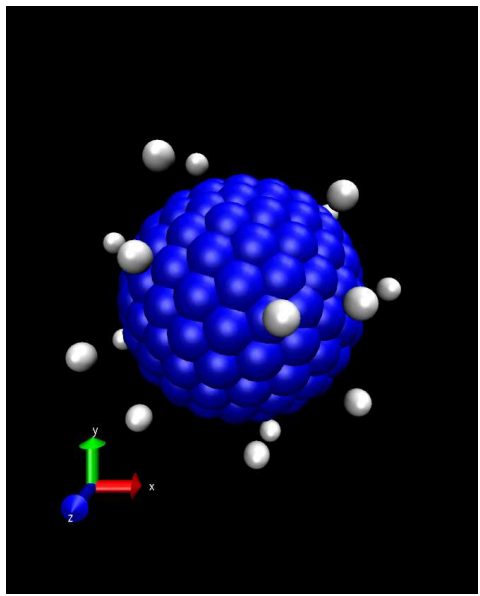}}
  \caption{Snapshot of a colloidal particle in a salt-free solution. The
surface sites are represented by the blue beads, and the white beads are
counterions. Solvent beads are not shown here. }
  \label{fig:snap}
\end{figure}

Specifically, we consider neutral colloids ($Q=0$) and colloids with positive
charge $Q=+50e$ in salt-free solutions and in salt solutions with salt ion
number density $\rho_s = 0.1 \sigma^{-3}$.  In these systems, the diffusion
constant of single microions is smaller than that of solvent particles
and concentration dependent, i.e., 
  $D_I=0.59 \pm 0.02 \,\sigma \sqrt{\varepsilon/m}$ 
in systems containing only counterions, and 
  $D_I=0.47 \pm 0.01 \,\sigma \sqrt{\varepsilon/m}$ 
in systems containing counterions and salt. Figure \ref{fig:snap} shows a 
representative snapshot of a single charged colloidal particle in salt-free 
solution.  

Our simulations were carried out using the open source package ESPResSo
\cite{ESPResSo}, with a slight modification that allows us to incorporate a
time-dependent external electric field.  A cubic simulation box of linear size 
$L=20 \sigma$ with periodic boundary conditions was used for all simulations.
Electrostatic interactions were calculated using the Particle-Particle-Particle
Mesh (P3M) method \cite{HockneyEastwood}, and a time step of $\Delta t = 0.01
\sigma \sqrt{m/\varepsilon}$ was used for the integration.

\begin{figure}[htbp]
  \centerline{\includegraphics[width=0.6\columnwidth]{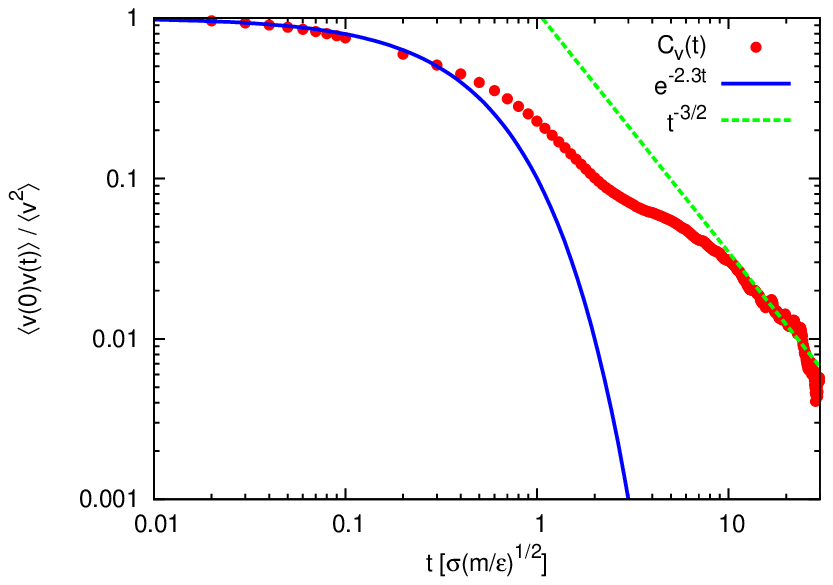}}
  \centerline{\includegraphics[width=0.6\columnwidth]{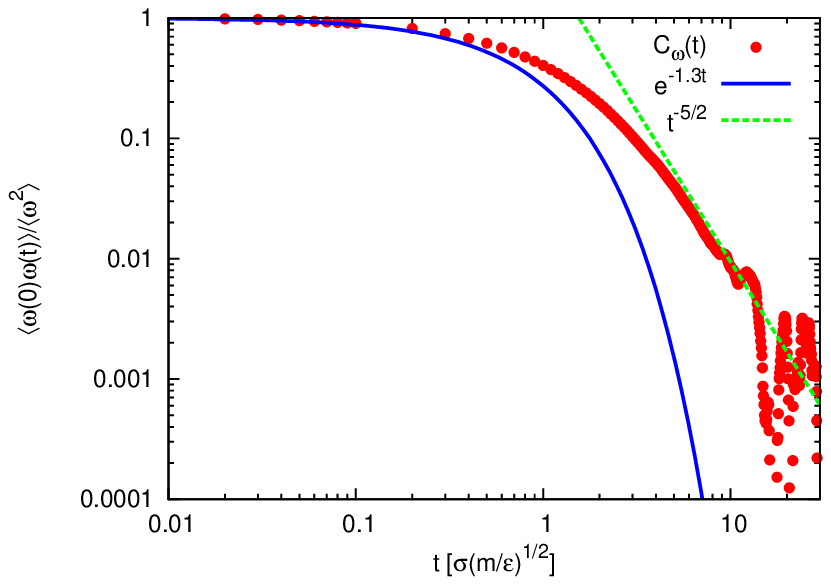}}
  \caption{Translational (top) and rotational (bottom) velocity autocorrelation
functions. The measurement is done for a single uncharged colloid with radius
$R=3.0\sigma$ in a salt-free solution. The temperature is $k_BT=1.0 \varepsilon$.} 
  \label{fig:acf}
\end{figure}

As a test of our colloidal model, we have performed simulations of an uncharged
colloid in a salt-free solution and measured the autocorrelation functions.  Two
autocorrelation functions were obtained from the simulations: the translational and
rotational velocity autocorrelation functions
\begin{eqnarray}
  \label{eq:vacf}
  C_v(t) &=& \frac{ \langle {\bf v}(0) \cdot {\bf v}(t) \rangle } 
  { \langle {\bf v}^2 \rangle }, \\
  \label{eq:wacf}
  C_{\omega}(t) &=& \frac{ \langle {\boldsymbol \omega}(0) \cdot {\boldsymbol \omega}(t) \rangle } 
  { \langle {\boldsymbol \omega}^2 \rangle }, 
\end{eqnarray}
where ${\bf v}(t)$ and ${\boldsymbol \omega}(t)$ are the translational velocity
and rotational velocity for the particle at time $t$, respectively. Figure
\ref{fig:acf} shows the simulation results averaged over ten runs with
different random number generator initializations. 

For short time lags, both autocorrelation functions show exponential relaxation.
The decay rate can be calculated using the Enskog dense-gas kinetic theory
\cite{Subramanian1975, Hynes1977}.  They are plotted as blue lines in Figure
\ref{fig:acf} and show reasonable agreement with the simulation data when
$t<0.1 \sigma \sqrt{m/\varepsilon}$.  Similar results have been obtained using
MPCD simulations \cite{Padding2005}.

For long time lags, hydrodynamic effects set in and lead to a slow relaxation
for autocorrelation functions of the colloidal particle \cite{Alder1970}. This
so-called long-time tail is the manifestation of momentum conservation, as the
momentum must be transported away from the colloids in a diffusive manner.
Mode-coupling theory predicts an algebraic $t^{-3/2}$ behavior at long times
for the translational velocity and $t^{-5/2}$ for the rotational velocity
\cite{HansenMcDonald}. The simulation results are plotted as green lines in
Figure \ref{fig:acf}. The data are consistent with the theoretical prediction
for $t>10 \sigma \sqrt{m/\varepsilon}$, but the rotational autocorrelation function
exhibits large fluctuations for large times.  This is mainly due to the fact
that the statistics for long time values becomes very bad, and very long
simulations are required in order to obtain accurate values.

From the velocity autocorrelation function, the diffusion constant of the
colloids can be calculated using the Green-Kubo relation
\begin{equation}
  D = \frac{1}{3} \int^{\infty}_0 dt \langle {\bf v}(0) \cdot {\bf v}(t) \rangle. 
\end{equation}
In our case, this gives the diffusion constant $D=0.013  \pm 0.002  \sigma
\sqrt{\varepsilon/m}$, which  is in good agreement with the theoretical value for
the diffusion constant of a Stokes sphere, $D=k_BT/(6\pi \eta_s R) = 0.0144 \pm
0.0002 \sigma \sqrt{\varepsilon/m}$.  

\section{Results and Discussion}
\label{sec:results}

In this section, we report our results for single colloidal particles of radius
$R=3.0\sigma$ under alternating electric fields.

\subsection{Displacement and velocity of the colloid}
\label{sec:pos_vel}

We start with a system containing one colloidal particle of charge $Q=+50 e$,
an equal amount of counterions with charge $-e$, and monovalent salt ions with
a number density $0.1 \sigma^{-3}$, resulting in a total of 450 negative
and 400 positive microions in the solution.

\begin{figure}[htbp]
  \centerline{\includegraphics[width=0.8\columnwidth]{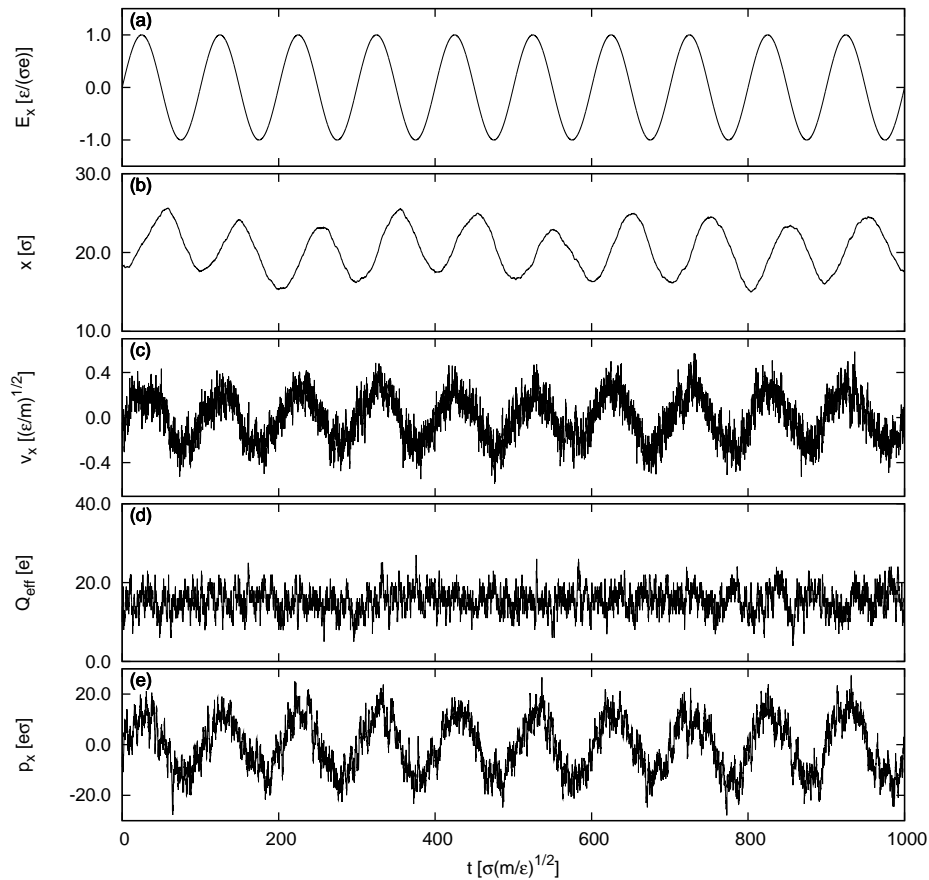}}
  \caption{Dynamics of a charged colloidal particle with charge $Q=+50e$ and
radius $R=3.0\sigma$ in a salt solution (number density of salt ions $0.1
\sigma^{-3}$) under the influence of an AC field with frequency $f = 0.01
\sqrt{\varepsilon/m}/ \sigma$ and amplitude $E_0 = 1.0 \varepsilon / (\sigma e)$. (a)
Instantaneous electric field. (b) Position of the particle along the
field direction. (c) Translational velocity along the field direction. (d)
Effective charge by counting number of ions inside the radius $r<4.0
\sigma$. (e) Dipole moment in the field direction. The effective charge and
the dipole moment are discussed in section \protect\ref{sec:ion_dip}.}
  \label{fig:time}
\end{figure}

Figure \ref{fig:time} shows the colloidal response to an AC field in
$x$-direction with frequency $f  =  0.01 \sqrt{\varepsilon/m} / \sigma$ and
amplitude $E_0 = 1.0 \varepsilon/(\sigma e)$.  After the AC field is turned on,
the system is allowed to evolve for $10^6$ time steps.  Then the measurement is
done every 10 time steps over a period of $10^6$ time steps.  In Figure
\ref{fig:time}, the time evolution of the electric field (top panel, Figure
\ref{fig:time}(a)) is contrasted with that of various quantities of interest
(Figure \ref{fig:time} (b)-(e)). We will begin with discussing the displacement 
and velocity, shown in Figure \ref{fig:time} (b) and (c). 

The electric field strength in this particular simulation was chosen relatively
large, such that the variations of the position and velocity are above the
thermal noise.  The period of the applied field is $T=1/f=100 \sigma
\sqrt{m/\varepsilon}$, which is less than the diffusion time scale of the
colloidal particle ($\tau^{(C)}_D = R^2/D \approx 640 \sigma
\sqrt{m/\varepsilon}$, where $D=0.013 \sigma \sqrt{\varepsilon/m}$ is the
diffusion constant of the colloidal particle). On the one hand, the particle thus
experiences many oscillation before diffusing over the distance of its radius.
On the other hand, the period is still much larger than the diffusion time
scale of the fluid ($\tau^{(S)}_D=\sigma^2/D_S \approx 1.5 \sigma
\sqrt{m/\varepsilon}$, where $D_S=0.66 \sigma \sqrt{m/\varepsilon}$ is the
diffusion constant of solvent particles). 

To rationalize the results, one may thus adopt a simplified picture where the
colloidal particle is viewed as a charged sphere immersed in a viscous fluid in
an AC-field. The equation of motion can be written as
\begin{equation}
  \label{eq:eom}
  M \ddot{x} + \gamma \dot{x} - QE_0 e^{i\omega t} = 0.
\end{equation}
The friction term is inversely proportional to the velocity and its value can be
approximated using the Stokes friction $\gamma=6\pi \eta_s R \approx 70
\sqrt{m\varepsilon}/\sigma$.  We solve the differential equation (\ref{eq:eom}) by
assuming that the position takes the form $x=A e^{i\omega t}$, where $A$ is a
complex number that characterizes the magnitude and phase of the oscillation.  The
results are
\begin{equation}
  |A|=\frac{QE_0}{\sqrt{M^2\omega^4 + \gamma^2 \omega^2}}, 
   \qquad \mathrm{Arg}(A) = \tan^{-1}(\frac{\gamma}{M\omega}) - \frac{\pi}{2}.  
\end{equation}
Inserting the parameters of our system, we obtain that the amplitude of the
position oscillation should be about $11 \sigma$, and the phase should be
advanced by about $\pi/2$ compared to the applied field.  Also, the amplitude
of the velocity oscillation should be $0.71 \sqrt{\varepsilon/m}$ and the
velocity should be almost in phase with the applied field.

The simulation results shown in Figure \ref{fig:time} are in good qualitative
agreement with the simplified theory.  The position variation clearly exhibits
a $\pi/2$ phase shift ahead of the applied field, while the velocity profile is
in phase.  However, the theory overestimates the amplitude of the oscillation.
The actual amplitude of the position oscillation is about $5 \sigma$, and that
of the velocity is about $0.3 \sqrt{\varepsilon/m}$.  One reason for this
discrepancy is that counterions condense on the colloids surface, which
effectively reduces the total charge of the colloids.  The binding of the
counterions and the colloid also increases the effective mass of the complex,
which further reduces the amplitude of the oscillations. Another effect comes
from the no-slip boundary condition of the colloidal particle: The fluid
particles adjacent to the colloid surface are dragged along with the colloid,
effectively increasing its mass.

\begin{figure}[htbp]
  \centerline{\includegraphics[width=0.8\columnwidth]{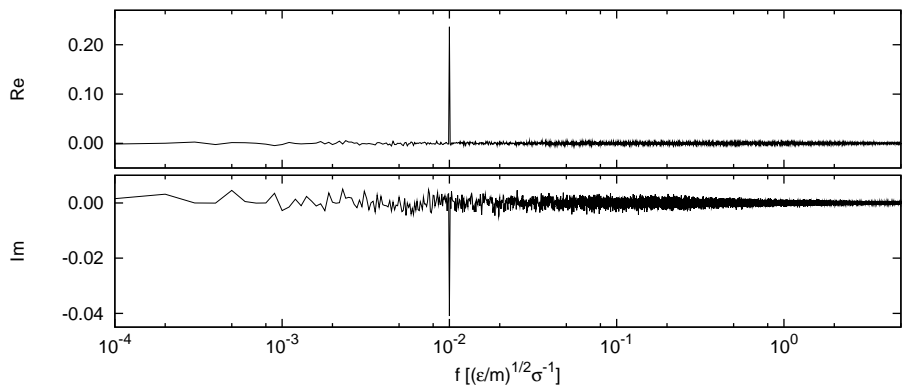}}
  \caption{Fourier transform of the time sequence of the colloid velocity 
of Figure \protect\ref{fig:time}(c).  The top panel shows the real part
corresponding to the in-phase component of the velocity, and the bottom panel
is the imaginary part corresponding to the out-of-phase component. Both the
real and the imaginary parts feature peaks at $f=0.01
\sqrt{\varepsilon/m}/\sigma$, the frequency of the external field.} 
  \label{fig:E1_vel_dft}
\end{figure}

One important quantity that characterizes the colloidal response to AC-fields
is the mobility $\mu(\omega)$.  In the presence of an external field 
${\bf E} e^{i\omega t}$, the particle oscillates with the velocity 
${\bf U} e^{i\omega t}$ with
\begin{equation}
  {\bf U} = \mu(\omega) {\bf E},
\end{equation}
where the mobility $\mu(\omega)$ is a complex function of the frequency.  In
the simulation, the time sequence of the velocity is noisy due to the thermal
fluctuation.  To obtain a clear signal and extract the mobility, we apply
Fourier transform to the time sequence of the velocity.  Figure
\ref{fig:E1_vel_dft} shows the Fourier transform of the time sequence in Figure
\ref{fig:time}(c).  The real and imaginary parts correspond to the in-phase and
out-of-phase components of the signal with respect to the applied field.  They
both exhibit peaks at the frequency of the external electric field, and the
peak values divided by the electric field strength give the real and imaginary
part of the mobility at that frequency. 

\subsection{Charge distribution and dipole moment}
\label{sec:ion_dip}

\begin{figure}[htbp]
  \centerline{\includegraphics[width=0.6\columnwidth]{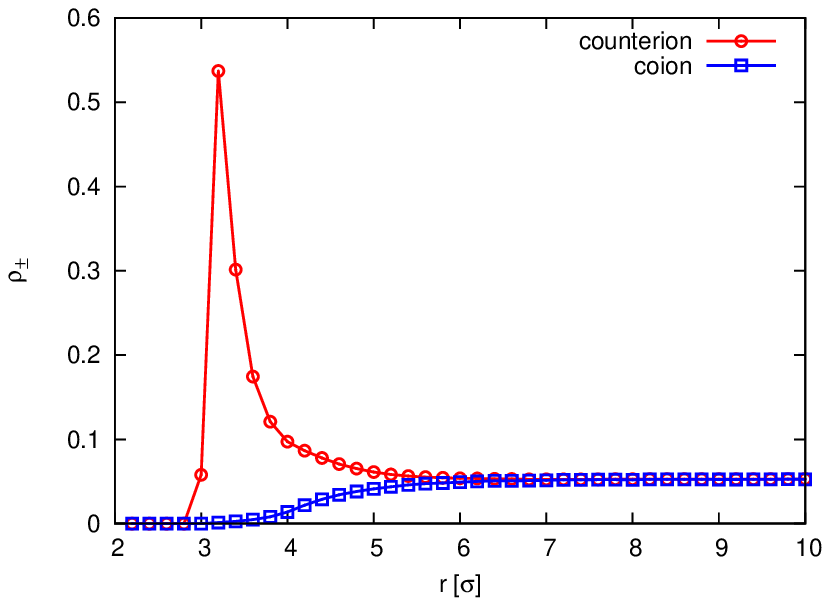}}
  \centerline{\includegraphics[width=0.6\columnwidth]{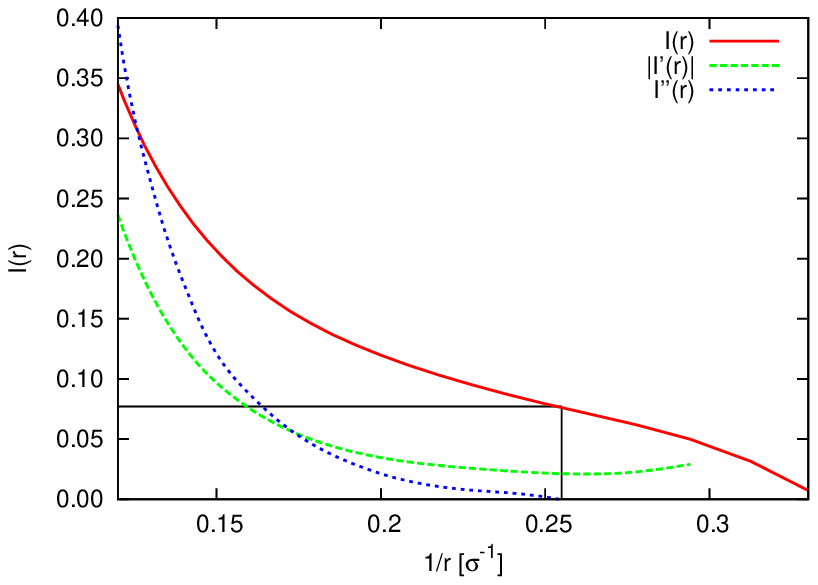}}
  \caption{(a) Average number density of counterions and coions as a function
of the distance to the colloid center. The average has been taken over $10^6$
time steps.  (b) The integrated counterion distribution $I(r)$ used to
determine the cutoff radius for the effective charge. The inflection point is
identified as the zero of the second derivative of $I(r)$ with respect to
$1/r$.} 
  \label{fig:E1_ion}
\end{figure}

Apart from allowing one to study the displacement and velocity of the colloidal
particle, the simulations also provide valuable information about the electric
double layer surrounding the colloid. During the simulation, we have collected
the histograms of counterion and coion distances to the center of the colloidal
particle.  Figure \ref{fig:E1_ion}(a) shows the time-averaged charge
distribution around the colloidal particle.  Due to the strong Coulomb
interaction with the positively charged colloids, counterions accumulate close
to the colloidal surface while the coions are depleted in the same region.  The
counterion concentration is peaked around $r=3.2 \sigma$ from the colloid
center.  Beyond that distance, the counterion concentration decays rapidly to
the bulk value, which is reached at around $r=6.0 \sigma$. At the same time,
coions are barely present within the distance $r<4.0 \sigma$, and the
concentration slowly increases to the bulk value, which is also reached at
around $r=6.0 \sigma$. 

The ion cloud surrounding a charged colloidal particle is commonly grouped into
two regions separated by a ``plane of shear'', which is defined as a hypothetical
plane that separates an immobilized liquid layer and the mobile fluid.  In the
region inside the shear plane, also called the stagnant layer, the ions are
strongly bound to the surface and have limited mobility.  Due to the friction
between the liquid and the particle surface, ions inside the stagnant layer
follow the surface.  Outside the stagnant layer, the ions are relatively free
to move.  The outer region up to the point where the ion concentration reaches
the bulk level is called diffuse layer. The thickness of the diffuse layer is
roughly determined by the Debye screening length, although it may effectively be
smaller due to nonlinear effects already at moderate surface charges
\cite{Smiatek2010}. The Debye screening length is given by
\begin{equation}
	\label{eq:Debye}
	l_D=\left[ 4\pi l_B \sum_i z_i^2 n_i(\infty) \right]^{-\frac{1}{2}},
\end{equation}
where $z_i$ and $n_i(\infty)$ are $i$-th ion's valence and bulk number density,
respectively, and the summation is over all ion species. In our system,
the Debye screening length is given by $l_D = 0.89 \sigma$ in pure solution
(neglecting the small contribution of the counterions).

Determining the extension of the stagnant layer is important because it
contains the ions that contribute to the effective charge of the particle.  In
simulations, however, it is difficult to pin down the exact position of the
shear plane.  Different methods have been proposed in the literature
\cite{Belloni1998, Deserno2000, Grass2009, Alexander1984, Lobaskin2004a}.  One
popular method is the inflection criterion \cite{Belloni1998, Deserno2000,
Grass2009}.  Here one first computes the integrated counterion distribution
$I(r)$ by counting the number of counterions inside a radius $r$ with respect
to the colloid center,
\begin{equation}
  I(r) = \frac{1}{N_-} \int_0^r \rho_- (r) dV, 
\end{equation}
where $\rho_-(r)$ is the counterion number density and the normalization
constant $N_-$ is the total number of counterions.  The value of $I(r)$
represents the fraction of counterions which are within a distance $r$ from the
colloid center.  The cutoff radius for summing up the effective charges is then
set at the inflection point (the point where the second derivative is zero) of
the integrated counterion distribution plotted versus the logarithmic distance
to the center. Although this criterion has originally been devised for the
static case, here we will attempt to use it for the dynamic situation by
applying it to the averaged counterion distribution. Figure \ref{fig:E1_ion}(b)
shows the integrated counterion distribution. The inflection point is
identified at $r=3.9 \sigma$. 

\begin{figure}[htbp]
  \centerline{\includegraphics[width=0.7\columnwidth]{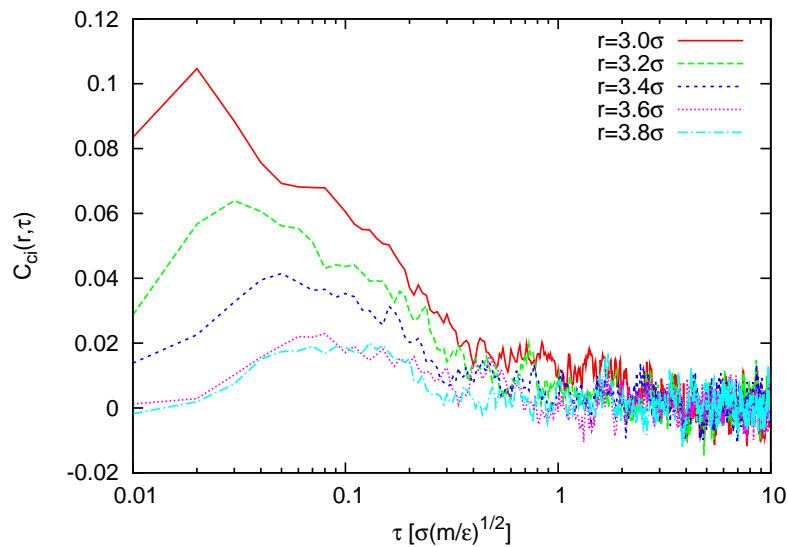}}
  \caption{Colloid-ion velocity correlation function $C_{ci}(r,\tau)$. The
values are averaged over two directions normal to the external field.}
  \label{fig:ccf}
\end{figure}

Another method to determine the shear plane is to compute the colloid-ion
velocity correlation \cite{Alexander1984, Lobaskin2004a}
\begin{equation}
  C_{ci}(r,\tau) = \langle v_c (t) v_i (r, t+\tau) \rangle
   - \langle v_c(t) \rangle \: \langle v_i(r,t+ \tau) \rangle ,
\end{equation}
where $v_c(t)$ is the velocity of the colloidal particle and $v_i(r,t)$ is the
velocity of ions which is located at $r$ with respect to the colloid center.
Since the motion of the colloid and the ions are strongly perturbed in the
direction of the external field, we only consider the correlation in the other
two directions. Figure \ref{fig:ccf} shows the correlation functions for five
different $r$. For ions which are located in the nearest surroundings of the
colloid surface, the correlation in the limit of $\tau \rightarrow 0$ is
positive, indicating that those ions are moving along with the colloid.  At larger
distance, the ion motion becomes anti-correlated with that of the colloid, and
the correlation $C_{ci}(\tau \rightarrow 0)$ turns negative.  The position
where the correlation first changes sign can also be viewed as a definition of
the shear plane. Using this method, one locates the shear plane at $r=3.8 \sigma$,
which is close to the value determined with the inflection criterion.

In order to characterize the dielectric response of the colloidal particle
under external fields, we must calculate the dipole moment of the system.
Whereas only the ions inside the stagnant layer contribute to the effective
charge, the dipole moment also has significant contributions from the diffuse
layer. Figure \ref{fig:E1_ion}(a) indicates that the ion concentrations reach
their bulk values at approximately $r=6.0\sigma$, therefore we computed the dipole
moment using a cutoff of $6.0\sigma$ from the colloid center.  The time
sequence of the effective charge and the dipole moment are plotted in Figure
\ref{fig:time}(d) and (e), respectively.  The averaged effective charge is
roughly $+15e$, corresponding to about condensed 35 counterions on the
colloidal surface.  The dipole moment oscillates with the same frequency
as the applied field.  The complex polarizability $\alpha(\omega)$ is obtained
by applying a Fourier transform to the time sequence of the dipole moment,
similar to the complex mobility discussed in last section.

\subsection{Influence of the strength of AC fields}
\label{sec:strength}

\begin{figure}[htbp]
  \centerline{\includegraphics[width=0.6\columnwidth]{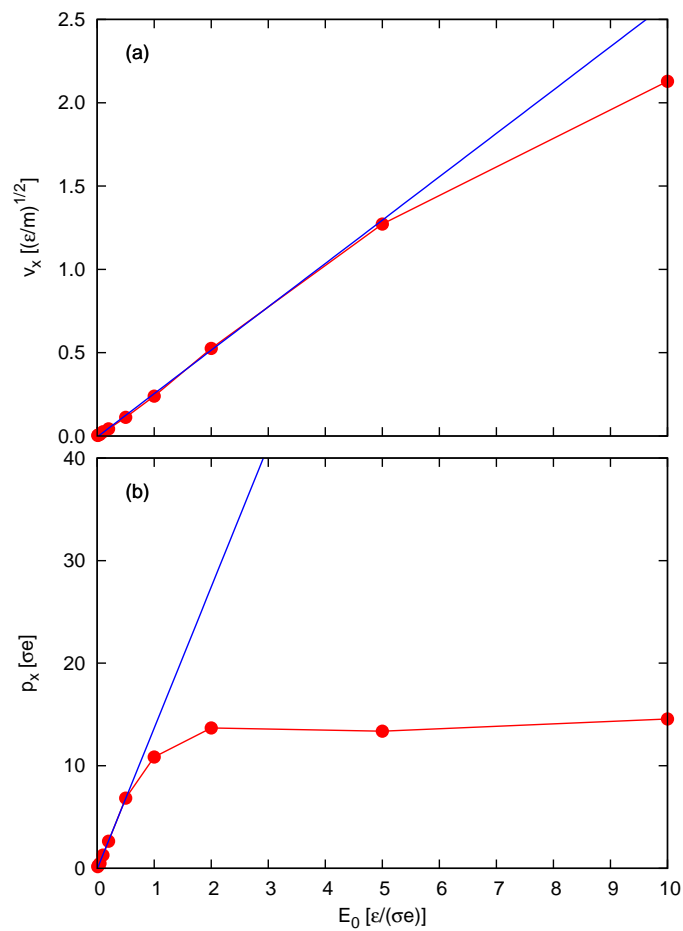}}
  \caption{(a) The amplitude of the colloid velocity as a function of the field
strength. (b) The amplitude of the dipole moment as a function of the field
strength. The frequency of the applied field is set at $f=0.01
\sqrt{\varepsilon/m}/\sigma$.}
  \label{fig:linear}
\end{figure}

Most theories on the dielectric response of charged colloids are based on
the linear response theory and assume the external electric field to be weak.
In our simulations, we can apply large electric fields to examine the effect
of the field strength and possible nonlinear effects. In this subsection, the
frequency is set at $f=0.01 \sqrt{\varepsilon/m}/\sigma$, which is in the
low-frequency regime (see next section).  The field strength is varied from
$0.01$ to $10 \varepsilon/(\sigma e)$.  By covering a wide range of electric field
amplitudes, we are able to estimate the limit of validity of the linear
response theory.  Figure \ref{fig:linear} shows the amplitudes of the colloid
velocity and dipole moment as a function of the field strength. 

\begin{figure}[htbp]
  \centerline{\includegraphics[width=0.8\columnwidth]{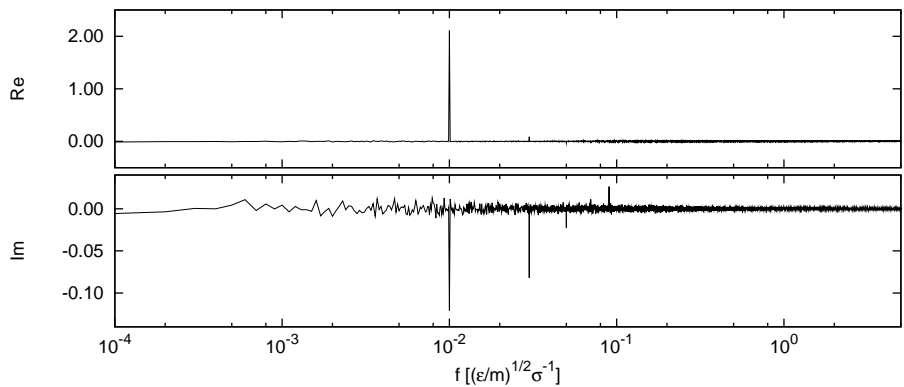}}
  \caption{Fourier transform of the time sequence of the colloid velocity for
a strong external field with frequency $f=0.01 \sqrt{\varepsilon/m} / \sigma$ and
amplitude $E_0=10 \varepsilon / (\sigma e)$. }
  \label{fig:E10_vel_dft}
\end{figure}

In weak fields, the velocity grows linearly as a function of the field strength
up to $E_0 \approx 2.0 \varepsilon / (\sigma e)$, corresponding to a constant
mobility.  For higher field strength, the growth slows down, which implies that
the mobility decreases.  This is different from the case of constant electric
field, where the mobility is found to increase above the linear region
\cite{Lobaskin2004a}.  Alternating electric fields of large field strength can
excite colloidal motion with signatures of various frequencies besides the
frequency of the applied field. This can be seen from the Fourier transform
of the colloid velocity, shown in Figure \ref{fig:E10_vel_dft}, where peaks at
higher frequencies appear for high field strength $E_0=10 \varepsilon / (\sigma
e)$ (see Figure \ref{fig:E1_vel_dft} for comparison). 

\begin{figure}[htbp]
  \centerline{\includegraphics[width=0.7\columnwidth]{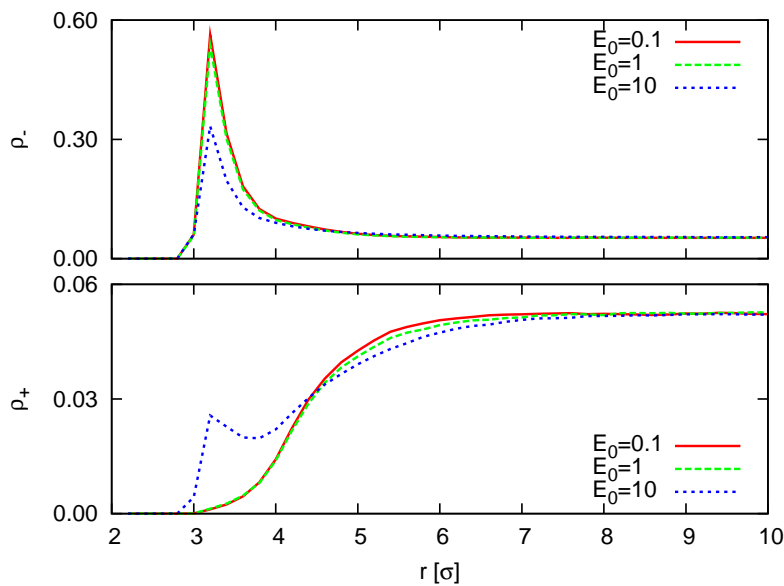}}
  \caption{Comparison of the ion distribution for three different field
strength $E_0 = 0.1, 1$ and $10 \varepsilon / (\sigma e)$. The top and bottom
figures are for the counterion and coion distribution, respectively.}

  \label{fig:ion}
\end{figure}

Looking at the dipole moment, one finds that the deviation from the linear
regime sets in at about $E_0 \approx 1.0 \varepsilon / (\sigma e)$.  Compared to
the colloid velocity, the deviation from the linear dependence is much more
pronounced for the dipole moment.  This is due to the fact that the dipole
moment measures how ions are distributed around the colloidal particle,
and the electric double layer can easily be influenced by the external field.
For intermediate field strength, the main effect of the external field is to
deform the shape of the electrical double layer, as the ions in the diffuse
layer are relatively mobile.  The shape of the ion clouds is elongated in the
field direction.  Figure \ref{fig:ion} shows the ion distribution for three
different field strength, $E_0 = 0.1$, $1$ and $10 \varepsilon / (\sigma e)$.  In
the weak field region ($E_0=0.1$ to $1.0 \varepsilon / (\sigma e)$), the change of
ion distribution is small.  When the field strength increases to $10 \varepsilon /
(\sigma e)$, the deviation becomes notable.  The peak of the counterion
distribution is reduced, and a second peak of coions close to the colloid 
surface develops.

When the field is strong enough, it may disturb the ions inside the stagnant
layer. Some of the ions originally bound to the colloidal surface will be
ripped off at large enough electric field.  This is demonstrated in Figure
\ref{fig:ion} where the counterion peak is reduced at higher field strength.
The required field strength can be estimated by equating the force exerted by
the external field and the electrostatic force from the charged colloids.
\begin{equation}
	E_0 = \frac{ Q_{\rm eff} }{ 4 \pi  \epsilon_m R^2 e} \approx 1.7 \varepsilon / (\sigma e),
\end{equation}
where we have used $Q_{\rm eff} = 15e$.  This number seems to determine the
point where the dipole moment as a function of field strength starts to level
off, indicating the breakdown of the weak field assumption.

\subsection{Frequency dependence}
\label{sec:frequency}

Finally, we investigate the effect of varying the frequency of the external AC
field.  The amplitude of the field is chosen in the linear region, $E_0 = 0.5
\varepsilon / (\sigma e)$.  The frequency is varied from $f=10^{-3}$ to $2.0
\sqrt{\varepsilon/m} / \sigma$.  A separate simulation with a constant electric
field is also performed to provide the reference at low-frequency limit.  As in
most previous simulations, the solution has a salt ion number density of $0.1
\sigma^{-3}$, corresponding to the Debye length $l_D = 0.89 \sigma$ in a system
without counterions and $l_D = 0.87 \sigma$ in a system with 50 counterions.
Thus the Debye length is smaller than, but comparable to the colloidal radius
($R=3.0 \sigma$).  We compare the simulations with the prediction of the
Maxwell-Wagner-O'Konski (MWO) theory, which is briefly sketched in
\ref{app:eMWO}.  Even though this theory was originally developed for
situations where the electric double layer is much thinner than the colloid, we
shall see that it still captures the most important features of the frequency
dependence on a semi-quantitative level. 

\begin{figure}[htbp]
 \centerline{\includegraphics[width=0.7\columnwidth]{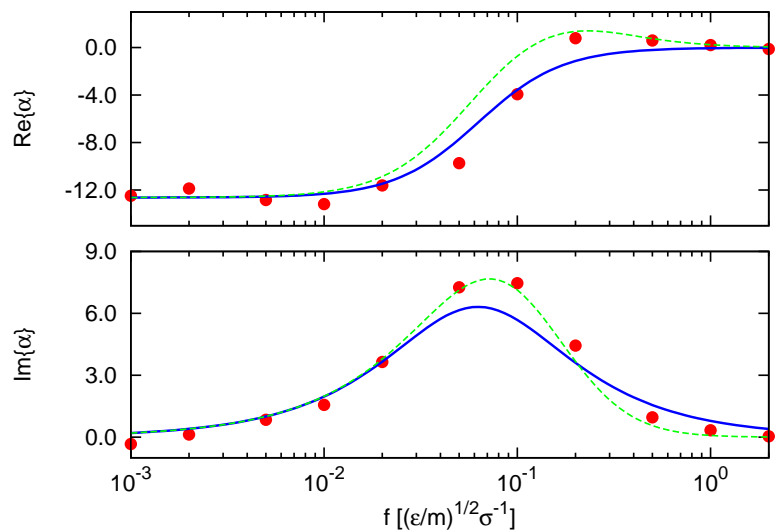}}
  \caption{Real and imaginary part of the complex polarizability of an
uncharged particle as a function of the frequency of applied electric field.
The field strength is set in the linear region $E_0=0.5 \varepsilon / (\sigma
e)$. The number density of the salt ions in the solution is $0.1 \sigma^{-3}$.
The points are simulation results. The solid lines give the prediction from the
Maxwell-Wagner theory with the effective colloid radius 
$R_{\mbox{\tiny eff}}=2.93 \sigma$. The dashed line shows revised prediction 
by taking into account of inertia effects. See text for explanation.}
  \label{fig:Q0_dip}
\end{figure}

We start with the easiest case of an uncharged colloidal particle, which has no
electric double layer.  Figure \ref{fig:Q0_dip} shows the polarizability for an
uncharged colloidal particle in salt solution with ions number density of
$0.1\sigma^{-3}$.  The real part of the polarizability shows the in-phase
component of the dipole moment with respect to the external field, while the
imaginary part gives the out-of-phase contribution.  The polarizability in the
low-frequency limit has a negative real part, indicating that the induced
dipole is anti-parallel to the applied field.  Uncharged colloids are not
surrounded by an electric double layer, and the dipole moment is induced solely
by the motion of the salt ions.  The positive ions move in the direction of
the electric field, and tend to accumulate on the back side of the colloid,
while the negative ions accumulate on the front end of the colloid.  This leads
to an effective dipole moment which points in the opposite direction than the
external field, sometimes referred to as volume polarization \cite{Dhont2010}. 

The solid line in Figure \ref{fig:Q0_dip} shows the prediction from the
Maxwell-Wagner theory.  The limiting behavior of the polarizability at
frequency $f \to 0$ was obtained by an independent run in a DC field and used
to calibrate the effective colloidal radius $R$ {\em via} $\alpha(0) = 4 \pi
\epsilon_m K_0 R^3$, which reduces to $\alpha(0) = -R^3/2$ in our system (see
\ref{app:eMWO}). We obtain $R_{\mbox{\tiny eff}} = 2.93 \pm 0.07 \sigma$, which
is reasonably close to the ``physical'' radius $R = 3.0 \sigma$.  The main
sources of discrepancy are the soft character of the excluded volume
interactions and the small size of the colloids. Having set the effective
colloidal radius, we can calculate the Maxwell-Wagner prediction without
further fit parameter, and the resulting curve is in quite good agreement with
the simulation data (Figure \ref{fig:Q0_dip}, solid line). Most notably, the
theory predicts a crossover between two regimes which is recovered in the
simulation at roughly the predicted crossover frequency. The simulations
feature a slight overshoot of the polarizability in the crossover region, which
can be explained by considering the inertia effect of the microions, which are
not entirely negligible in our simulations.  (In real systems, they are
negligible.)  When microions with mass are immersed in a viscous fluid, their
response to the external electric fields depends on the frequency.  Thus the
conductivity becomes frequency-dependent.  The inertia effect can be
incorporated into the theory (see \ref{app:eMWO}) and the theory can be revised
accordingly, giving the dashed line in Figure \ref{fig:Q0_dip}. It predicts
small overshoots in both the real and the imaginary part of the polarizability,
which are consistent with the simulation. 

\begin{figure}[htbp]
  \centerline{\includegraphics[width=0.7\columnwidth]{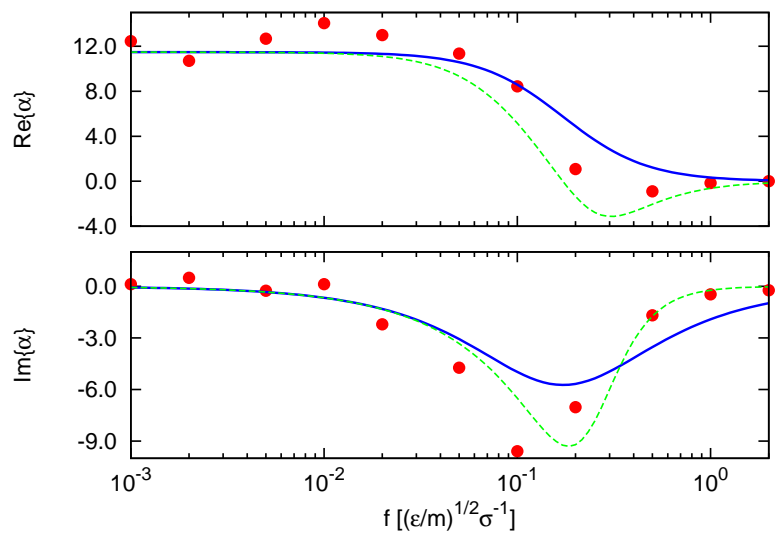}}
  \caption{Same as Figure \protect\ref{fig:Q0_dip} for a charged colloid
($Q=50e$). The points are simulation results. The solid lines give the
prediction from the MWO theory, using an effective colloid size $R_{\mbox{\tiny
eff}}=2.93\,\sigma$ and the surface conductance $K_{\sigma}=0.24 
e^2/(\sigma\sqrt{m\varepsilon})$. The dashed line shows revised prediction by
taking into account of ions' inertia effect. See text for explanation.}
\label{fig:Q50_dip}
\end{figure}

The situation is more complicated when the colloids are charged, due to the
presence of the electric double layer.  Figure \ref{fig:Q50_dip} shows the
polarizability for a charged colloid ($Q=50e$) in salt solution (same salt
concentration as before).  

In the low-frequency region $f<10^{-2} \sqrt{\varepsilon/m} / \sigma$, the
external perturbation is slow enough that the system can follow. The dipole
moment is in phase with the applied field, as reflected by the fact that ${\rm
Im}\{\alpha\}$ is close to zero. The real part ${\rm Re}\{\alpha\}$ is
positive, in contrast to that obtained for uncharged colloids, indicating that
the dipole moment points in the same direction as the applied field.  In the
absence of external fields, the center of the counterion cloud and the center
of the spherical particle coincide, and the dipole moment is zero. If one
applies an electric field, the positively charged particle moves in the
direction of the field, while the negative ion cloud moves in the opposite
direction.  The resulting net dipole moment points in the same direction as the
external field, therefore one has ${\rm Re}\{\alpha\}>0$ in the low-frequency
limit.  In the opposite limit of high frequency, the colloid and the ion cloud
can no longer follow the field, thus both ${\rm Re}\{\alpha\}$ and ${\rm
Im}\{\alpha\}$ converge to zero. At intermediate frequencies $f \sim 0.1
(\varepsilon/m)^{1/2} \sigma^{-1}$, the real part ${\rm Re}\{\alpha\}$ crosses
over from positive to zero with an overshoot below the transition frequency and
a slight undershoot to negative values after the transition frequency.  The
imaginary part ${\rm Im}\{\alpha\}$ assumes a maximum, indicating that the
response is out of phase and that there is high dissipation.

According to the Maxwell-Wagner-O'Konski (MWO) theory, the effect of the
electric double layer can be taken into account by introducing a surface
conductance in the Maxwell-Wagner theory. Here we have determined the value of
the surface conductance by calculating the $f\to 0$ limit of the polarizability
from an independent simulation in a DC field, assuming that the effective
radius is the same for charged and uncharged colloids.  The simulation gives
the surface conductance $K_{\sigma}=0.24 \pm 0.07
e^2/(\sigma\sqrt{m\varepsilon})$, which is consistent within the error with the
theoretical estimate based on the extended MWO theory \cite{Saville2000} (see
appendix), $K_{\sigma}=0.19 e^2/(\sigma\sqrt{m\varepsilon})$. Using this value,
we can again calculate the theoretical polarizability as a function of
frequency without any fit parameter. The resulting curves without and with ion
inertia effects are shown in Figure \ref{fig:Q50_dip} with solid and dashed
lines, respectively.  The prediction of the MWO theory is in surprisingly good
agreement with the simulations, given how questionable its validity is.  The
theory captures the main qualitative features, and roughly the correct
crossover frequency.  Similar to the uncharged case, the dielectric response at
high frequency are affected by the finite mass of microions.  The revised
theory shows better agreement with the simulation after the transition
frequency.  The overshoot in the real part ${\rm Re}\{\alpha\}$ may be an
indication of the low-frequency dielectric dispersion \cite{DukhinShilov}, and
further studies are required to clarify this problem.

\section{Summary}
\label{sec:summary}

We have carried out mesoscopic molecular dynamics simulations for a colloidal
particle under alternating electric fields.  We have accounted in full for the
hydrodynamic and electrostatic interactions as well as thermal fluctuations,
using Dissipative Particle Dynamics and a Particle-Particle-Particle Mesh
method for evaluating Coulomb interactions.
 
We obtained information about the displacement and velocity of the colloidal
particle.  A simple classical theory is proposed to explain the simulation
results at low frequency and shows positive agreement.  We have also considered
the contribution from the electric double layer surrounding the colloid.  Two
different methods, namely, the inflection criterion and the colloid-ion
correlation function, were used to determine the position of the shear plane,
and it was shown that they both give the same results.  Furthermore, the dipole
moment was calculated by integrating all contributions of ions from the colloid
surface to the point where the ion concentration reaches the bulk value. 

The effect of the external field strength was investigated systematically.  By
computing the mobility and the polarizability for various field strength, we
have examined the validity of the linear response theory.  Nonlinear effects
were found to set in for electric strength up to $1.0 \varepsilon/(\sigma e)$,
which is higher than the values reported for constant electric fields
\cite{Lobaskin2004a, Chatterji2005, Duenweg2008}.  This can be contributed to
the excitation of high-frequency modes and the stripping of the counterions
that are bounded to the colloid surface.

Furthermore, the frequency dependence of the polarizability was studied for an
uncharged colloid and a charged colloid in a salt solution.  The simulation
results were compared with the predictions of the Maxwell-Wagner-O'Konski
theory.  Despite the fact that thermal fluctuations are substantial and that
the size of the colloids is comparable to the Debye screening length, the MWO
theory performed surprisingly well, especially for uncharged colloids.  The
inertia effect of microions can also be incorporated into the theory to explain
the high frequency behavior. For charged colloids, the theory and simulations
show noticeable deviations, indicating that the thin assumption of a thin
electrostatic double layer is questionable for our system. 


\ack{The authors would like to thank Burkhard D\"unweg, Roman Schmitz, Christian Holm, Axel Arnold, Olaf Lenz, Jens Smiatek, Peter Virnau, Shuanhu Qi and Stefan Medina Hernando. 
This work was funded by the Deutsche Forschungsgemeinschaft (DFG) through the SFB-TR6 ``Physics of Colloidal Dispersions in External Fields''.  Computational resources at John von Neumann Institute for Computing (NIC J\"ulich), High Performance Computing Center Stuttgart (HLRS) and Mainz University are gratefully acknowledged.}

 \appendix

\section{Maxwell-Wagner-O'Konski theory}
\label{app:eMWO}

In this appendix, we give a short introduction to the Maxwell-Wagner-O'Konski
theory. More detailed information can be found in Refs. \cite{Maxwell1954,
Wagner1914,OKonski1960,Saville2000}.

When loss is present, the dipole moment of a particle immersed in fluids
exhibits a phase lag with respect to the external sinusoidal field. A complex
effective dipole moment can be written as $  \mathbf{p} = 4 \pi \epsilon_m
K(\omega) R^3 \: \mathbf{E}$ (cf. Equation (\ref{eq:mosotti})), where the
Clausius-Mossotti factor $K(\omega)$ is a complex number containing both the
magnitude and the phase information about the effective dipole moment.  
In the Maxwell-Wagner theory, it has the form
\begin{equation}
  \label{eq:cm2}
  K (\epsilon^*_p, \epsilon^*_m) 
    = \frac{ \epsilon^*_p - \epsilon^*_m }{ \epsilon^*_p + 2\epsilon^*_m }, 
\end{equation}
where $\epsilon^*_p$ and $\epsilon^*_m$ are the complex dielectric constants of
the particle and the medium, respectively.  They are defined as
\begin{equation}
  \epsilon^*_p = \epsilon_p + \frac{ K_p }{i\omega}, \quad
  \epsilon^*_m = \epsilon_m + \frac{ K_m }{i\omega},
\end{equation}
where $\epsilon$ (without the star) and $K$ are the permittivity and conductivity, respectively. 
For $\epsilon_m=\epsilon_p=\epsilon$, the Clausius-Mossotti factor can be rewritten as
\begin{equation}
\label{eq:K}
K =K_0 \: \frac{1-i \tilde{\omega}}{1+\tilde{\omega}^2}
\end{equation}
with
\begin{equation}
K_0 = \frac{K_p - K_m}{K_p + 2 K_m} \quad \mbox{and} \quad
\tilde{\omega} = \frac{3 \epsilon}{K_p + 2 K_m} \: \omega.
\end{equation}
In our system, we have $\epsilon_p = \epsilon_m = 1/4 \pi$, $K_p = 0$, thus
$K_0 = -1/2$ is negative. The conductivity
of the medium $K_m$ is related to the microion diffusion constant $D_I$  {\em via}
$K_m = n_{\rm ion} e^2 D_I / (k_B T)$ for 1-1 electrolytes.
 
The classical Maxwell-Wagner theory fails to explain the dielectrophoretic
properties of latex particles.  Latex has a low intrinsic conductivity, but the
measurement indicated that the particle conductivity is high.  It was concluded
that the electric double layer surface significantly contributes to
the particle conductivity.  This was first demonstrated by O'Konski
\cite{OKonski1960}
\begin{equation}
  \label{eq:OKonski}
  K_p \rightarrow K_p + \frac {2K_{\sigma}}{R}
\end{equation}
where $K_{\sigma}$ is the surface conductance (unit S instead of
S$\cdot$m$^{-1}$ for conductivity) due to the electrostatic double layer.  The
extra contribution to the conductivity increases the Maxwell-Wagner relaxation
frequency without introducing an extra dispersion (no extra extrema in the
imaginary part of the polarizability).  The idea is to encapsulate the
contribution from the ion cloud around the colloids into one single parameter
(the surface conductance).  This is justified for high ionic strength, highly
charged colloids and high frequencies. Another important effect of $K_{\sigma}$
is to shift the prefactor $K_0$ in Eq.\ (\ref{eq:K}) towards positive values,
to the point that the dielectric response changes sign if $K_{\sigma}$ is
sufficiently large.

We still need a way to obtain the surface conductance $K_{\sigma}$ from the 
effective surface charge density $\sigma$ in the simulation.  
This can be done by the procedure outlined in \cite{Saville2000}. 
First, the $\zeta$-potential is solved using the following formula \cite{RSS},
\begin{equation}
  \sigma = \frac{ \epsilon_m k_BT }{e l_D} \left[ 2 \sinh \left( \frac{e\zeta}{2k_BT} \right) + \frac{4}{R/l_D} \tanh \left( \frac{e\zeta}{4k_BT} \right) \right].
\end{equation}  
The surface conductance then can be related to the $\zeta$-potential by
Bikerman's expression
\begin{equation}
  \label{eq:Bikerman}
  K_{\sigma} = l_D \left[ \exp (| \frac{e \zeta}{2k_BT} |) -1 \right] (1+3m) K_m, 
\end{equation}
where $m$ is a dimensionless ionic drag coefficient \cite{Bikerman1940,OBrien1986}.
Bikerman's expression is applicable for 1-1 electrolytes, and the
contribution from coions has been neglected. 
The ionic drag coefficient is related to the viscosity $\eta_s$ and ion diffusion constant $D_I$,
\begin{equation}
  m = \frac{ 2\epsilon_m (k_B T)^2 } { 3 \eta_s e^2 D_I },
\end{equation}
($m=0.092$ for our system, and $m=0.18$ for KCl).  
With $Q=50e$ and $R_{\rm eff}=2.93\sigma$, we obtain the surface conductance $K_{\sigma}=0.19\, e^2/(\sigma\sqrt{m\varepsilon})$.

The inertia effect of ions can be taken into account by considering a charged particle with charge $e$, mass $m$ immersed in a viscous fluid. 
The equation of motion for the particle under an AC field $E_0 e^{i\omega t}$ is
\begin{equation}
  m \ddot{x} = - \gamma \dot{x} + eE_0 e^{i\omega t}, 
\end{equation}
where $\gamma = k_BT/D_I$ is the friction constant. 
After solving the equation of motion, one finds that the velocity of the particle has the form
\begin{equation}
  \label{eq:v_omega}
  \dot{x} \propto \frac{ 1 - i ( \frac{m\omega}{\gamma} )}{ 1 + (\frac{m\omega}{\gamma})^2}. 
\end{equation}
Since the conductivity of salt solutions is proportional to the ion's velocity, it is reasonable to assume that the conductivity has the same frequency dependency
\begin{equation}
  \label{eq:Km_omega}
  K_m \rightarrow K_m \frac{ 1 - i ( \frac{m\omega}{\gamma} )}{ 1 + (\frac{m\omega}{\gamma})^2}. 
\end{equation}
The dashed lines in Figure \ref{fig:Q0_dip} and \ref{fig:Q50_dip} are obtained by substituting (\ref{eq:Km_omega}) into the Clausius-Mossotti factor (\ref{eq:K}).

\section*{References}
\bibliography{ac}

\end{document}